# Magnetic relaxation and correlating effective magnetic moment with particle size distribution in maghemite nanoparticles

K. Pisane[1], E. Despeaux[2], and M. S. Seehra[1]*

[1]Department of Physics & Astronomy, and [2]Department of Pharmaceutical Sciences

West Virginia University, Morgantown, WV 26506 USA

## ABSTRACT

The role of particle size distribution inherently present in magnetic nanoparticles (NPs) is examined in considerable detail in relation to the measured magnetic properties of oleic acid-coated maghemite ($\gamma$-$Fe_2O_3$) NPs. Transmission electron microscopy (TEM) of the sol-gel synthesized $\gamma$-$Fe_2O_3$ NPs showed a log-normal distribution of sizes with average diameter $\langle D \rangle$ = 7.04 nm and standard deviation $\sigma$ = 0.78 nm. Magnetization, M, vs. temperature (2 K to 350 K) of the NPs was measured in an applied magnetic field H up to 90 kOe along with the temperature dependence of the ac susceptibilities, $\chi'$ and $\chi''$, at various frequencies, $f_m$, from 10 Hz to 10 kHz. From the shift of the blocking temperature from $T_B$ = 35 K at 10 Hz to $T_B$ = 48 K at 10 kHz, the absence of any significant interparticle interaction is inferred and the relaxation frequency $f_o$ = 2.6 x $10^{10}$ Hz and anisotropy constant $K_a$ = 5.48 x $10^5$ ergs/cm$^3$ are determined. For $T < T_B$, the coercivity $H_C$ is practically negligible. For $T > T_B$, the data of M vs. H up to 90 kOe at several temperatures are analyzed two different ways: (i) in terms of the modified Langevin function yielding an average magnetic moment per particle $\mu_p$ = 7300(500) $\mu_B$; and (ii) in terms of log-normal distribution of moments yielding $\langle \mu \rangle$ = 6670 $\mu_B$ at 150 K decreasing to $\langle \mu \rangle$ = 6100 $\mu_B$ at 300 K with standard deviations $\sigma \simeq \langle \mu \rangle / 2$. The above two approaches yield consistent and physically meaningful results as long as the width parameter, s, of the log-normal distribution is less than 0.83.

**KEY WORDS:** Maghemite nanoparticles; Magnetic properties; Magnetic relaxation; Particle size distribution; Magnetic moment distribution; Interparticle interaction.

## 1. Introduction:

Magnetic properties of nanoparticles (NPs) depend not only on their size, size distribution and morphology but also on several other factors such as interparticle interactions, interactions between the surface spins and ligands, the presence of defects, and the degree of ordering in the surface and core spins and interactions between them. Understanding these different effects requires detailed investigations of the magnetic properties as a function of temperature, magnetic fields, and measuring frequencies on structurally well-characterized samples of different sizes, which are often synthesized by different methods. The interesting physics emanating from such investigations and diverse applications of magnetic NPs in catalysis, ferro-fluids, high-density magnetic storage and biomedicine have continued to attract the interest of researchers worldwide [1-6].

Nanoparticles of magnetite ($Fe_3O_4$) and maghemite ($\gamma$-$Fe_2O_3$), the two stable oxides of iron with ferrimagnetic ordering, have potential applications in ferrofluids and in biomedicine (targeted drug delivery, and magnetic hyperthermia) [1,4-6]. Bulk $Fe_3O_4$ ($\gamma$-$Fe_2O_3$) has a spinel structure with lattice constant a= 0.839 nm (0.835 nm), ferrimagnetic Néel temperature $T_{FN} \simeq$ 948 K (600 K) and 8 (32/3) formula units per cell with the following cationic arrangements on the A and B sites: 8 $Fe_3O_4$= 8 $[Fe^{3+}]_A$ $[Fe^{3+} \cdot Fe^{2+}]_B$ $O_4$ and (32/3) $\gamma$-$Fe_2O_3$ =[8 $Fe^{3+}]_A$ [ (40/3) $Fe^{3+} \cdot (8/3)$ V] $O_{32}$ where V represents a vacancy. For $Fe_3O_4$, the ferrimagnetic moment results from $Fe^{2+}$ ions, leading to calculated saturated magnetization $M_S$ = 106 emu/g using $\mu(Fe^{2+})$ = 4.4 $\mu_B$ in the limit of T → 0 K. A similar calculation for $\gamma$-$Fe_2O_3$, in which ferrimagnetism results from unequal numbers of $Fe^{3+}$ ions on the A and B sites, yields $M_S$ = 87 emu/g using $\mu(Fe^{3+})$ = 5 $\mu_B$. Because of the high magnitudes of $T_{FN}$, the measured magnitudes of $M_S$ at room temperature are expected to be only slightly lower than the above values. These high magnetization values for $Fe_3O_4$ and $\gamma$-$Fe_2O_3$, combined with their resistance to oxidation and biocompatibility, make them highly suitable for biomedical applications [1,4-6]. The often-observed decrease of room temperature super-paramagnetic magnetization with decrease in particle size, D, in both $Fe_3O_4$ [7] and $\gamma$-$Fe_2O_3$ [8], has been interpreted in terms of a core-shell structure with a magnetically dead shell of about 1 nm not contributing to $M_S$. However, in $\gamma$-$Fe_2O_3$ NPs with inherent presence of cationic vacancies, spin disorder in the core spins has also been concluded from the observations of large coercivity $H_C$ and large high-field susceptibility [9,10]. Thus, in addition to size, the method of preparation of the $\gamma$-$Fe_2O_3$ NPs may also determine their magnetic properties. The recent studies by Vaishnava et al. [11] on several different size NPs of $\gamma$-$Fe_2O_3$ dispersed in a polystyrene resin matrix have addressed the issue of the size distribution on the measured dc magnetic properties.

In this paper, we report results of our investigations of the dc and ac magnetic properties of $\gamma$-$Fe_2O_3$ NPs of size D $\simeq$ 7 nm, covering the temperature range from 2 K to 350 K in dc magnetic fields H up to 90 kOe and in ac measuring frequencies $f_m$ from 10 Hz to 10 kHz. Measurements of the ac susceptibilities as a function of $f_m$ are essential for determining the relaxation rate needed for magnetic hyperthermia applications and for determining the strength of any interparticle interactions [12,13]. In our previous studies of $\gamma$-$Fe_2O_3$ NPs [14] and those of Vaishnava et al. [11], the ac susceptibility studies were not carried out. In this paper, analysis of the magnetic data on the 7 nm $\gamma$-$Fe_2O_3$ NPs prepared by the method of Hyeon et al. [15] and structurally characterized by x-ray diffraction (XRD), transmission electron microscopy (TEM), thermo-gravimetric analysis (TGA) and diffuse reflectance spectroscopy (DRS) is reported with the following important results: (i) The blocking temperature $T_B \simeq 35$ K is determined from dc magnetization measurements carried out under the ZFC (zero-field-cooled) and FC (field-cooled) conditions; (ii) From the analysis of change in $T_B$ with measuring frequency $f_m$, negligible interparticle interaction is inferred for these oleic acid (OA)-coated particles with the relaxation attempt frequency $f_o \simeq 2.6 \times 10^{10}$ Hz of the Néel-Brown relaxation [16,17]; (iii) For $T > T_B$, the fit of the M vs. H data to the modified Langevin function yields the magnetic moment per particle $\mu_p \simeq 7300$ $\mu_B$; (iv) Measurements by TEM yield a log-normal size distribution of NPs with average diameter $\langle D \rangle = 7.04$ nm and standard deviation $\sigma = 0.78$ nm; (v) The effects of the log-normal particle size distribution on the particle moment distribution are analyzed and correlated with the results obtained from the modified Langevin variation fit; (vi) and for $T < T_B$, the negligible measured coercivity $H_C \approx 15$ Oe, suggests coherent rotation of the spins under applied H and well-ordered spins in the nanoparticles. Details of the experimental procedures, experimental results, and their analysis and discussion are presented below.

## 2. Theoretical Considerations

Here we briefly summarize the basic equations in NP magnetism which are used later for the interpretation and discussion of the experimental results. For non-interacting NPs with volume V and anisotropy $K_a$, the frequency $f$ for the reversal of its magnetic moment $\mu_p$ is given by the Neel-Brown relaxation [16.17]:

$$f = f_o \exp(-T_a/T) \quad (1).$$

Here $T_a = K_a V/k_B$ where $k_B$ is the Boltzmann constant and $f_o$ is the attempt frequency. The blocking temperature $T_B$ defined by $f = f_m$ (measuring frequency) is then given by

$$T_B = T_a / \ln(f_o/f_m). \quad (2)$$

In the presence of interparticle interactions (IPI), Eq. (2) is replaced by the Vogel-Fulcher relaxation equation [18, 19] leading to

$$T_B = T_o + [T_a / \ln(f_o/f_m)] \quad (3)$$

where $T_o$ is a measure of the strength of the IPI. For uncoated, strongly interacting NiO NPs, $T_o$= 70 K was determined with $f_o$= 9.2 x $10^{11}$ Hz [20]. According to Eq. 3, IPI effectively increase the energy barrier of the system, a conclusion also reached by the theoretical/computational studies of Chantrell et al. [21]. The presence of IPI in NPs can also be determined from the measured variation of $T_B$ with change in $f_m$ by calculating the quantity $\Phi$ given by [22]

$$\Phi = \Delta T_B/T_B \Delta log_{10} f_m \qquad (4).$$

It is known [22] that $\Phi \geq 0.13$ for non-interacting particles and $\Phi$ is very small (0.005-0.05) for spin glasses. The intermediate values of $\Phi$ (0.05 $\leq \Phi \leq$ 0.13) signify the presence of IPI with decreasing effect for increasing values of $\Phi$. The magnitude of $\Phi$ can be determined from the change in $T_B$ measured using ac susceptibilities $\chi'$ and $\chi''$ as a function of temperature at a frequency $f_m$, since $\chi''$ peaks at $T_B$ and $\chi''=C\partial(T\chi')/\partial T$ [12,23].

For T > $T_B$, the NPs are in the super-paramagnetic state and M vs. H data are expected to obey the modified Langevin function [24, 25]:

$$M = M_o \mathcal{L}(\mu_p H/k_B T) + \chi_a H \qquad (5).$$

Here $\mathcal{L}(x)=\coth(x)-(1/x)$ is the Langevin function, $M_o$ is the temperature dependent saturation magnetization, $\mu_p$ is the average magnetic moment per particle, and $\chi_a$ is the linear component of the susceptibility usually estimated from the high-field data. In the case of a size distribution of particles, Eq. 5 has to be modified to include a corresponding distribution of the magnetic moments $\mu_p$ of the particle [26, 11]. This issue is discussed later in the paper.

**3. Synthesis and Structural Characterization:**

To synthesize the γ-$Fe_2O_3$ NPs, the procedure described by Hyeon et al. [15] was employed since this method is known to produce highly crystalline NPs [14, 15]. Briefly, 0.2 mL of $Fe(CO)_5$ was added to a mixture of octyl ether (10 mL) and oleic acid (1.28 g) at 100 $^0$C and refluxed for one hour resulting in a change of color of the solution from orange to black. The solution was then allowed to cool to room temperature, 0.34 g of dehydrated trimethylamine N-oxide $(CH_3)_3NO$ was added and the solution was heated to 130 $^0C$ and held at that temperature for 2 hours. The temperature was then raised slowly until reflux, and held for one hour. The solution was then allowed to cool to room temperature, and ethanol was added until a black precipitate formed. A strong magnet was used to hold the precipitate in place as the excess liquid was poured off. The NPs were re-suspended in a small amount of toluene and ethanol was again used to precipitate the NPs. This washing process was repeated three times before the particles were stored in toluene until used.

The transmission electron micrograph (TEM) of the sample obtained using a JEOL JEM-2100 transmission electron microscope is shown in the inset of Fig. 1. Using 'image J' software, dimensions and areas of the NPs were measured. The aspect ratio of 87 % of the NPs was found to be less than 1.3 indicating nearly spherical shape. The particle diameters were calculated from the areas assuming spherical shapes; a histogram of this distribution is shown in Fig. 1. The size distribution is fit to the log-normal distribution:

$$f(D) = \frac{1}{2\pi\lambda_D D} exp\left\{\frac{-[\ln(D/D_o)]^2}{2\lambda_D^2}\right\} \qquad (6)$$

with $D_o$= median particle diameter and $\lambda_D$= width of the distribution with average particle diameter $\langle D\rangle=D_o exp(\lambda_D^2/2)$ and the standard deviation $\sigma=\langle D\rangle[exp(\lambda_D^2)-1]^{1/2}$. The fit shown by the solid line in Fig. 1 yields $\lambda_D$= 0.11, $D_o$= 7.0 nm, $\langle D\rangle$= 7.04 nm and $\sigma$= 0.78 nm.

The x-ray diffraction patterns of a commercial bulk $\gamma$-$Fe_2O_3$ sample and the synthesized 7 nm sample, measured using a Rigaku RU-300 x-ray Diffractometer with Cu $K_\alpha$ radiation ($\lambda$=

0.15418 nm), are shown in Fig. 2. For the bulk sample, the observation of the weak superstructure (210) and (211) lines is assigned to some ordering of the cationic vacancies on the B sites. In the inset of Fig. 2, the plot of $\beta\cos\theta$ vs. $\sin\theta$ according to the Williamson-Hall relation [14,27] for the 7 nm sample is shown where $\beta$ is the instrument corrected full-width at half-maximum (in radians) of the major Bragg lines and $\eta$ is the strain. This analysis yields the crystallite size = 5.2 nm and the strain $\eta = 4 \times 10^{-3}$.

The oleic acid (OA) used for stabilizing and surface coating the $\gamma$-$Fe_2O_3$ NPs has the formula $C_{18}H_{34}O_2$ with the bonding $(CH_3)(CH_2)_7CH=CH(CH_2)_7COOH$ and boiling point= 360 $^0$C. To verify the surface coating with OA, the IR spectra obtained by diffuse reflectance spectroscopy (Perkin-Elmer) is shown in Fig 3. The IR spectrum of the NPs contains broad peaks around 1644-1520 $cm^{-1}$ and 1461-1313 $cm^{-1}$ for the symmetric and asymmetric stretching modes of the carboxyl group, respectively. The possible values of $\Delta=\upsilon_a(COO^-)-\upsilon_s(COO^-)$ range from 59-331 $cm^{-1}$ and indicate the presence of unidentate ($\Delta>200$ $cm^{-1}$), bidentate ($\Delta<100$ $cm^{-1}$), and bridging (intermediate values of $\Delta$) complexes at the NP surface [28,29]. In the NP spectrum, the peak near 1710 $cm^{-1}$ due to the carbonyl stretching mode (C=O) in oleic acid is also present, albeit quite weak, indicating that some of the oleic acid is not bound covalently to the NPs but may be attached via hydrogen bonds [28, 29].

In order to accurately determine the mass of $\gamma$-$Fe_2O_3$ in the OA coated $\gamma$-$Fe_2O_3$, thermo-gravimetric analysis (TGA) using TA Instruments Model Q50 was done. The results are depicted in Fig. 4 in terms of the change of weight of the sample with temperature as the sample is heated at 5 $^o$C/minute in flowing $N_2$ gas. The few percent change near 100 $^0$C is due to loss of absorbed moisture. The weight loss rate beginning around 200 $^0$C and ending near 500 $^0$C with a peak near 360 $^0$C is associated with the evaporation of OA. The results of Fig. 4 show that nearly 80 % weight of the sample is due to $\gamma$-$Fe_2O_3$. This correction is used later in the normalization of the measured M (in emu) to magnetization per unit mass of $\gamma$-$Fe_2O_3$ in the magnetic measurements since OA is only weakly diamagnetic and any contribution from it to the measured M is comparatively negligible.

**4. Results from Magnetic Measurements:**

The measurements of M vs. H and T were done employing the ac measurement system of the Physical Property Measurement System (PPMS) purchased from Quantum Design Inc. The data shown here were corrected for the weak diamagnetic background signal of the sample holder. For the ZFC case, the sample was cooled to 2 K in H = 0 Oe, a measuring field H was then applied and M vs. T data taken with increasing temperature to 350 K after stabilizing the temperature at each point. For the field-cooled (FC) case, the sample is cooled to 2 K in a non-zero H and M vs. T data taken similarly with increasing temperatures.

The temperature dependence of $\chi$= M/H for the ZFC and FC modes in H=100 Oe in Fig. 5 shows the characteristic blocking temperature $T_B$= 35 K below which $\chi$(FC) bifurcates from $\chi$(ZFC). The broadness of the peak in the $\chi$(ZFC) vs. T data is likely due to the distribution of the particle sizes shown in Fig. 1 with the smaller (larger) particles getting unblocked at the lower (higher) temperatures. The plots of M vs H at 2 K for T < $T_B$ and for T > $T_B$ (T=100, 150, 200, 250 and 300 K) are shown in Fig. 6. The inset shows negligible hysteresis of the data even at 2 K. The temperature dependence of the measured coercivity $H_C$ for temperatures between 2 K and 35 K is shown in the other inset of Fig. 6. The measured $H_C$ < 15 Oe is within the uncertainty of setting H = 0 Oe so that $H_C$ is practically negligible. This is comparable to the observations reported by Dutta et al. [14] in their similarly prepared 7 nm sample of $\gamma$-$Fe_2O_3$.

However, a number of previous researchers [8-11] have reported kOe level $H_C$ values in samples of γ-$Fe_2O_3$ NPs prepared by other methods. This issue along with the measured values of $M_o$ being dependent on the preparation method and the samples size is discussed later.

The temperature dependence of the ac susceptibility $\chi'=M'/H_{ac}$ and $\chi''=M''/H_{ac}$ at the measuring frequencies $f_m$ from 10 Hz to 10 kHz using $H_{ac}$= 10 Oe and $H_{dc}$=0 Oe is shown in Fig. 7. Since the peaks in $\chi''$ and $\partial(T\chi')/\partial T$ represent $T_B$ [12,23], this variation of $T_B$ vs. $f_m$ is plotted in the inset of Fig. 8. The increase in $T_B$ with increasing $f_m$ is expected from Eq. 2. The determination of $f_o$ and $T_o$ from this variation is discussed below.

## 5. Analysis and Interpretation
### 5.1 Neel-Brown relaxation and interparticle interaction

First, the analysis of the variation of $T_B$ with the measuring frequency $f_m$ in zero dc field shown in the inset of Fig. 8 is presented. From Eq. (2), a plot of $1/T_B$ vs. $\ln(f_m)$ should yield a straight line with the slope being $(-1/T_a)$ and the intercept equal to $\ln(f_o)/T_a$. This analysis shown in Fig. 8 yields $T_a$= 725 ± 5 K and $f_o$~2.6x10$^{10}$ Hz. This magnitude of $f_o$ is close to the magnitude expected for the NPs of ferromagnetic materials in the absence of any IPI such as $f_o$ = 1.8 x 10$^{10}$ Hz observed in Ni NPs dispersed in $SiO_2$ [12]. For NPs of antiferromagnetic materials, the magnitude of $f_o$ is usually higher by an order of magnitude because of the enhanced effective anisotropy resulting from coupling between anisotropy and exchange energies; as examples $f_o$ = 5 x 10$^{12}$ Hz and 2 x 10$^{11}$ Hz have been reported respectively for NPs of antiferromagnetic NiO [20] and ferrihydrite [30,31].

The strength of the interparticle interaction can be estimated from the data of $T_B$ vs. $f_m$, by calculating Φ from Eq. 4 which yields Φ= 0.12, close to Φ= 0.13 expected for non-interacting

NPs. Thus, we infer that in these NPs, IPI are negligible likely due to the oleic acid coating on the particles. Therefore, $T_o= 0$ K of Eq. 3 is inferred.

## 5.2. Analysis of M vs. H data above $T_B$:

The analysis of the M/H vs. T data shown in Fig. 5 and M vs. H data at different T shown in Fig. 6 is considered next. For $T > T_B$, the data is often analyzed in terms of the modified Langevin relation of Eq. 5 with the three fitting parameters of $M_o$, $\mu_p$ and $\chi_a$. Using the three parameter least–squares fits to similar M vs. H data above $T_B$ in doped ferrihydrites [25] and in ferrihydrite nanoflakes [32], fits to the data were demonstrated with relatively consistent values of $\mu_p$ but $M_o$ and $\chi_a$ decreasing with increasing T as expected. Similar analysis of the data of Fig. 6 is shown in Fig. 9 as a plot of $(M - \chi_a H)/M_o$ vs H/T on a semi-log plot to highlight the good fit of the data for all H/T values. This fit yields $\mu_p = 7300(500)$ $\mu_B$ as the average magnetic moment per particle with the temperature dependence of $M_o$ and $\chi_a$ shown in the inset of Fig 9. The parameters determined from the data are given in Table 1. An extrapolation of $M_o$ to T= 0 K yields $M^* \approx 65$ emu/g. Using $\mu_p = M^*\rho V$ with $V = (\pi/6)D^3$ for spherical particles and density $\rho = 4.856$ g/cm$^3$ for maghemite, yields D= 7.5 nm. This magnitude of D is in good agreement with average $\langle D \rangle$ determined by TEM (Fig. 1).

In the limit of $(\mu_p H/k_B T) \ll 1$, Eq. 5 leads to [33]

$$\chi = \chi_o + (C/T), \quad C = \mu_p M^*/3k_B \quad (7)$$

where $\chi_o = \chi_a - (C/T_N)$. In Fig. 10, the plot of $\chi^{-1}$ vs. T using the data of Fig. 5 shows the non-linear behavior expected from Eq. 7. Plotting $\chi$ vs. (1/T) and taking the limit $(1/T) \to 0$ yields $\chi_o = 0.0186 \pm .0005$ emu/g Oe. Then, a plot of $(\chi-\chi_o)^{-1}$ vs. T does exhibit linear behavior for T >

$T_B$ as expected from Eq. 7, and yields C= 14.6 ± 0.1 emu K/g Oe. Using $\mu_p = 3k_B C/M^*$ from Eq. 7 yields $\mu_p$ = 10027 (140) $\mu_B$. This magnitude is considerably larger than $\mu_p$= 7300 $\mu_B$ determined earlier from the fit to Eq. 5. This discrepancy is due to the fact that in the low-field limit of Eq. 5 used to derive Eq. 7, the variation of M vs. H is dominated by contributions from the larger particles in the size distribution [34,35], thus resulting in a larger magnitude of derived $\mu_p$ determined from the fit to Eq. 5. The importance of this analysis is to demonstrate that in the super-paramagnetic regime, $\chi$ does not follow the Curie law but Eq. 7 and $\mu_p$ determined from such an analysis is dominated by contributions from the larger particles of the distribution in the particle size.

### 5.3 Analysis using distribution of magnetic moments

It is reasonable to expect a distribution in the magnitudes of $\mu_p$ if there is distribution in the particle sizes since $\mu_p = M^*\rho V$ as noted earlier. This issue was addressed in the earlier paper by Ibrahim et al. [34] based on the analysis by Richardson and Desai [35] and more recently by Silva et al. [26] in connection with the data on ferrihydrite NPs. Since then, it has been applied to the magnetic properties of NiO NPs [36] and $\gamma$-Fe$_2$O$_3$ NPs [11]. Analogous to the distribution in size given in Eq. 6, the distribution in magnetic moment $\mu$ is written as

$$f(\mu) = \frac{1}{\mu s\sqrt{2\pi}} \exp\left\{\frac{-[\ln(\mu/\mu_o)]^2}{2s^2}\right\} \quad (8)$$

where $\mu_o$ is the median value of $\mu$ and s describes the width of the distribution of particle moments in the Langevin fit so that Eq. 5 is replaced by

$$M = \frac{N}{s\sqrt{2\pi}} \int_0^\infty L(\mu H/k_B T) \exp\left\{\frac{-\left[\ln\left(\frac{\mu}{\mu_o}\right)\right]^2}{2s^2}\right\} d\mu + \chi_a H \quad (9).$$

Here, N is the number of particles per gram such that the product of N and the average particle moment described by $\langle\mu\rangle=\mu_o\exp(\frac{s^2}{2})$, gives the temperature-dependent saturation magnetization $M_o$. Applying Eq. 9 in place of Eq. 5, the M vs. H data were fit to determine the magnetic moment distribution parameters and the average particle magnetic moment as a function of temperature following the procedure of [11,26]. The parameters for this fit are given in Table 2. It is evident that the magnitudes of N and the width parameter s are practically temperature-independent as expected whereas $M_o$ decreases by few percent with increase in temperature. The magnitudes of $\langle\mu\rangle$ are slightly lower than the average $\mu_p$= 7300 (500) $\mu_B$ determined from the fit to the modified Langevin function of Eq.5. The fits of the M vs. H data to Eq. 9 at several temperatures using the parameters of Table 2 are shown in Fig. 11. The distribution functions $f(\mu)$ vs. $\mu$ derived from these fits for T = 150 K and 300 K are shown in the inset of Fig. 11.

## 5.4. Temperature dependence of magnetization below $T_B$

Since $T_B$ is directly proportional to the volume of the particles (Eq. 2), this will lead to a distribution of $T_B$ similar to the size distribution (Eq. 6) and can be written as [37, 11]

$$f(T_B)=\frac{1}{\sqrt{2\pi}T_B\lambda_B}\exp\left\{-\frac{\left[\ln(\frac{T_B}{T_{B0}})\right]^2}{2\lambda_B^2}\right\}. \quad (10)$$

Following [37,11], the variation of $f(T_B)$ with $T_B$ can be determined from the difference $\Delta M=M(FC)-M(ZFC)$ data of Fig. 5 and theoretically given by

$$\Delta M=\frac{20M_{sp}^2H}{3K_a\langle T_B\rangle}\int_T^\infty T_B f(T_B)dT_B \quad (11)$$

where $M_{sp}$ is the spontaneous magnetization at H= 100 Oe and $\langle T_B \rangle$ is the average blocking temperature. The value 20 in Eq. 10 comes from $\ln(f_o/f_m)-1$ using the value of $f_o$ determined earlier and $f_m = 20$ Hz for dc measurements for our magnetometer. The fit to Eq. 11 is shown in Fig. 12 with $M_{sp}$= 87.8 ± 0.7 emu/cm$^3$, H=100 Oe, $K_a$= 5.48 x 10$^5$ erg/cm$^3$, $T_{B0}$= 14 ± 0.4 K and $\lambda_B$= 0.37 ± 0.02. This value of $\lambda_B$ is in good agreement with the expected width in particle volumes $\lambda_v$= 0.33 determined from the measured particle diameters. Using $\langle T_B \rangle = T_{B0}\exp(\lambda_B^2/2)$ yields $\langle T_B \rangle$=15±0.5 K as the average blocking temperature. This issue $\langle T_B \rangle$ of being less than $T_B$= 35 K in a system with size distribution has been discussed [26, 30, 37] in that $T_B$ represents the maximum blocking temperature in H=100 Oe above which all particles of different sizes are unblocked. The plot of $f(T_B)$ vs. $T_B$ using these evaluated parameters is shown in the inset of Fig. 12.

## 6. Discussion:

From the analysis given above it is evident that size distribution often present in most real NP systems needs to be taken into account in the analysis of the magnetic data. The fit of the M vs. H data above $T_B$ to Eq. 5 has often been used to determine $\mu_p$ as done here in Fig. 9 and in several other papers [24-26, 32,36]. A comparison of the parameters in Tables 1 and 2 determined from the fits of the M vs. H data to Eq. 5 and Eq. 9 respectively shows meaningful agreement between $\mu_p$ and $M_o$ values determined from the fit to the modified Langevin function of Eq.5 and $\langle \mu \rangle$ and $M_o = N\langle \mu \rangle$ determined using the size distribution of Eq. 9. The important conclusion from this analysis is that this should always be possible as long as the width parameter of the log-normal distribution, s, is considerably smaller than 0.83. For the log-normal distribution, the average $\langle \mu \rangle = \mu_o \exp[s^2/2]$ with the standard deviation $\sigma = \langle \mu \rangle \left[ \exp(s^2) - 1 \right]^{1/2}$. From these expressions, it follows that $\sigma = \langle \mu \rangle$ for s=0.83 and for s > 0.83, σ becomes larger than $\langle \mu \rangle$, leading to unphysical comparison between $\mu_p$ of Eq. 5 and the quantities $\langle \mu \rangle$ and $\mu_o$ evaluated from Eq. 9. For example, in ferritin NPs, Silva et al. [26] found s = 0.9 (1.3) at 30 K (65 K) and in NiO NPs, Tiwari et al. [36] reported s = 1.30 (1.26) at 320 K (350 K). In such cases, the distribution in magnetic moments is not only due to size distribution of the NPs but there is also contribution from the uncompensated moments in the core of the NPs. This leads to severe disagreement between $\mu_p$ determined from Eq. 5 and $\langle \mu \rangle$ and $\mu_o$ evaluated from Eq. 9 as reported in the case of NPs of NiO [36] and ferritin [26]. For example, in NiO NPs at 320 K with s =1.30, the reported values are: $\mu_p$ = 1841 $\mu_B$ from fit to Eq. 5 and $\langle \mu \rangle$ = 305 $\mu_B$ with $\mu_o$ = 130 $\mu_B$ from fits to Eq. 9. Obviously, these magnitudes are vastly different and attempts to make these quantities of $\mu_p$, $\langle \mu \rangle$ and $\mu_o$ represent the same physical parameter are not appropriate. This will be true when s > 0.83. The noteworthy point here is that the magnitude of s > 0.83 implies that size distribution alone is not sufficient to explain the distribution of magnetic moments in the NPs. In the results reported here in Table 2 for the 7 nm γ-Fe$_2$O$_3$ NPs, the magnitude of s = 0.48 so that a good correlation observed between $\mu_p$, $\langle \mu \rangle$ and $\mu_o$ supports the arguments given above. In summary, the modified Langevin function of Eq. 5 may be applied successfully only if s is sufficiently less than 0.83.

## 7. Conclusions:

Experimental results and their analysis/interpretation on the magnetic properties of OA-coated 7 nm γ-Fe$_2$O$_3$ NPs have been presented with a special focus on the role of particle size distribution on the interpretation of the M vs. H data at different temperatures and M vs. T data at different measuring $f_m$. From the analysis of the change in T$_B$ with change in $f_m$, absence of any significant interparticle interaction is inferred, likely due to the oleic acid coating, similar to the observation in oleic acid-coated NiO compared with uncoated NP$_S$ [20, 38]. This analysis also yielded $f_o$ = 2.6 x $10^{10}$ Hz (the attempt frequency for the Neel-Brown relaxation) and anisotropy constant $K_a$ = 5.5 x $10^5$ ergs/cm$^3$. Another notable observation is the absence of any significant coercivity $H_C$ in this system, which is different from large $H_C$ reported in other γ-Fe$_2$O$_3$ NPs prepared by different methods [8-10]. Thus, the magnetic properties of γ-Fe$_2$O$_3$ NPs are dependent on the synthesis route. Finally, the effect of particle size distribution on the average magnetic moment per particle was analyzed first by the modified Langevin function without taking into consideration the size distribution and then by including size distribution. It is shown that these approaches give reasonably consistent results if the width of the moment distribution, s, is considerably less than 0.83. For larger values of s, the concept of an average magnetic moment is not meaningful and the fit of the data to Eq. 5 is not advisable.

**Acknowledgments**

This work was supported in part by a grant from the U.S. National Science Foundation (grant #DGE-1144676). We acknowledge use of the WVU Shared Research Facilities.

**Table 1**: Magnitudes of the various parameters of Eq. 5 obtained from the fit to the data shown in Fig. 9. $R^2$ provides a measure of the quality of the fit to the data with $R^2 = 1$ representing a perfect fit. The numbers in parenthesis are the estimated uncertainties.

| Temp (K) | $M_o$ (emu/g) | $\chi_a$ ($10^{-5}$ emu/g/Oe) | $\mu_p$ ($\mu_B$) | $R^2$ |
|---|---|---|---|---|
| 100 | 53.2 (0.4) | 4.42 (1.05) | 7836 (268) | 0.9925 |
| 150 | 51.6 (0.4) | 3.61 (1.07) | 7388 (151) | 0.9983 |
| 200 | 49.8 (0.4) | 3.56 (1.11) | 7290 (152) | 0.9984 |
| 250 | 50.7 (0.4) | 3.46 (0.98) | 7325 (131) | 0.9990 |
| 300 | 47.2 (0.3) | 3.40 (0.86) | 6823 (117) | 0.9993 |

**Table 2**: Magnitudes of the various parameters of Eq. 9 obtained from the fit to the data shown in Fig. 11. The numbers in parenthesis are the estimated uncertainties.

| Temp (K) | $N$ ($10^{17}$/g) | $\mu_o$ ($\mu_B$) | s | $\chi_a$ ($10^{-5}$ emu/g/Oe) | $M_o$ (emu/g) | $\langle \mu \rangle$ ($\mu_B$) | $R^2$ |
|---|---|---|---|---|---|---|---|
| 150 | 8.37 (0.10) | 5958 (5) | 0.475 (0.016) | 3.59 (0.64) | 51.8 (1.1) | 6670 (55) | 0.9990 |
| 200 | 8.37 (0.20) | 5750 (216) | 0.478 (0.036) | 3.59 (0.39) | 50.9 (4.0) | 6554 (364) | 0.9995 |
| 250 | 8.36 (0.08) | 5846 (5) | 0.480 (0.013) | 3.40 (0.53) | 50.9 (0.9) | 6560 (46) | 0.9992 |
| 300 | 8.35 (0.07) | 5463 (5) | 0.470 (0.011) | 3.37 (0.36) | 47.2 (0.7) | 6101 (36) | 0.9995 |

**Figures and Captions**

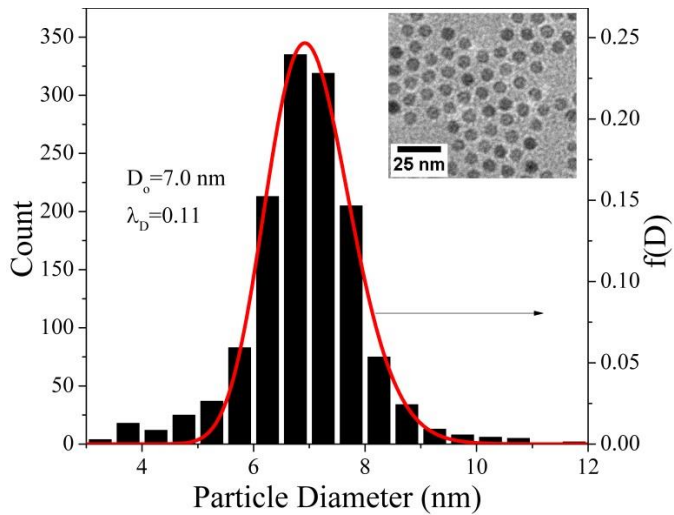

**Fig.1**: Histogram of the measured particle diameters as determined from average particle areas measured from TEM. The solid line is the fit to the log-normal size distribution f(D) of Eq. 7. The inset is a representative micrograph showing nearly spherical particles.

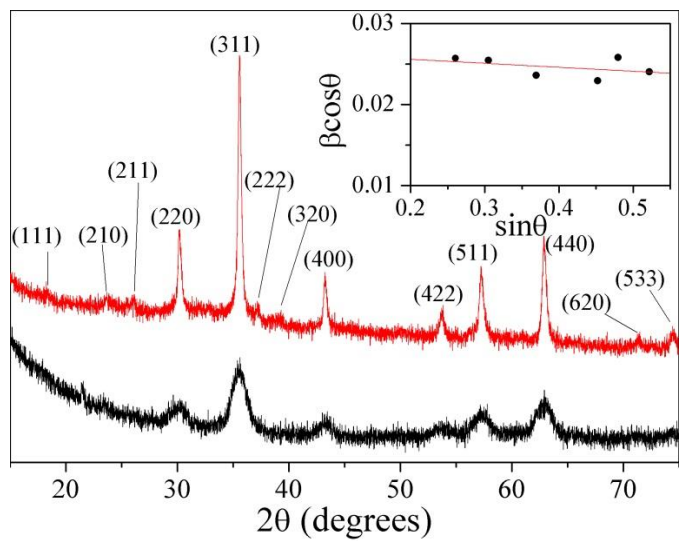

**Fig.2.** XRD patterns of the commercial bulk γ-Fe$_2$O$_3$ powder (top) and the γ-Fe$_2$O$_3$ NPs (bottom). The inset shows a plot of βcos(θ) vs. sin(θ) of the Williamson-Hall equation for the NPs with the slope of the linear fit yielding strain η= -5 x 10$^{-3}$.

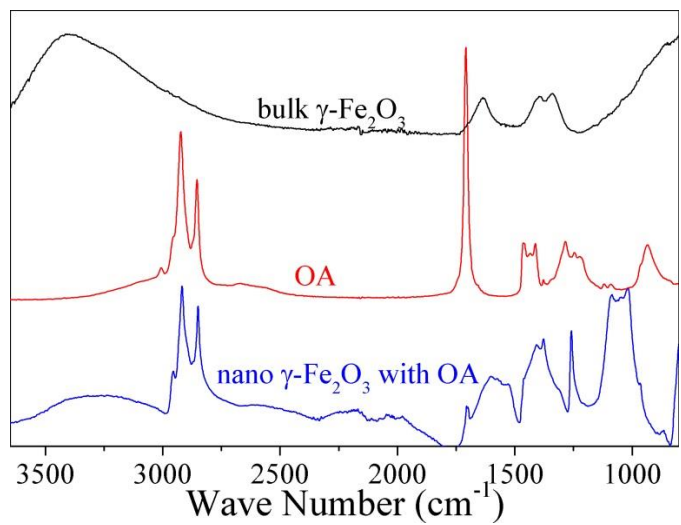

**Fig. 3**. FTIR spectra for bulk γ-Fe$_2$O$_3$ (top), pure oleic acid (middle), and OA-coated γ-Fe$_2$O$_3$ NPs (bottom).

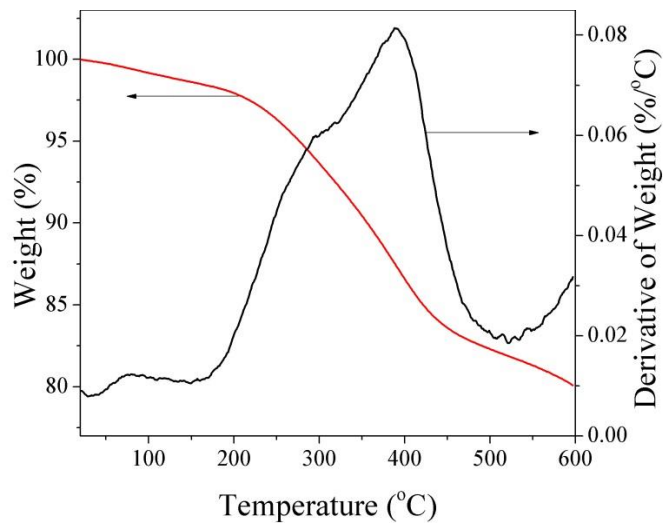

**Fig. 4**: Plots of % change in weight, W, with temperature, T, and computed $d$W/ $d$T for the oleic acid-coated NPs as the sample is heated at 5 °C per minute in flowing $N_2$ gas.

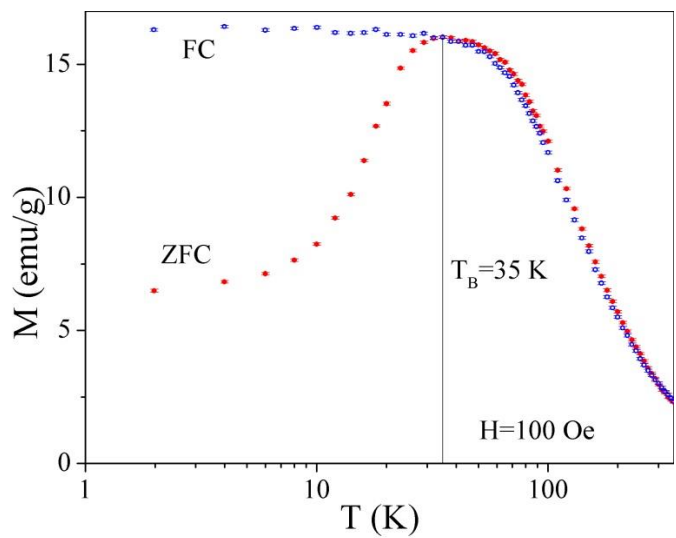

**Fig. 5**: Plots of the measured magnetization M vs. temperature for the ZFC (filled circles) and FC (open circles) conditions in H = 100 Oe. The bifurcation and peak in the ZFC data yields $T_B$ = 35 K. The experimental uncertainties in M are within the size of the data points.

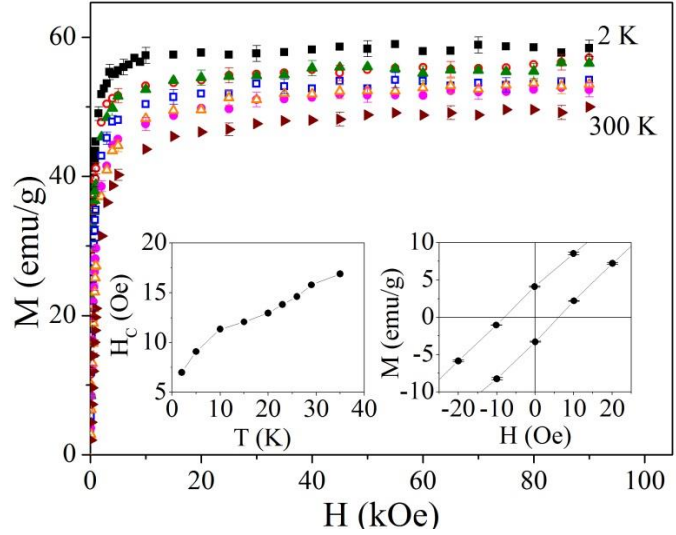

**Fig. 6**: Plots of M vs. H measured at T = 2, 50, 100, 150, 200, 250 and 300 K. The right inset shows the low field behavior of the hysteresis loop measured at 2 K with coercivity $H_C$ ~15 Oe and the left inset shows the temperature dependence of $H_C$. Error bars for select data points show typical uncertainties in M.

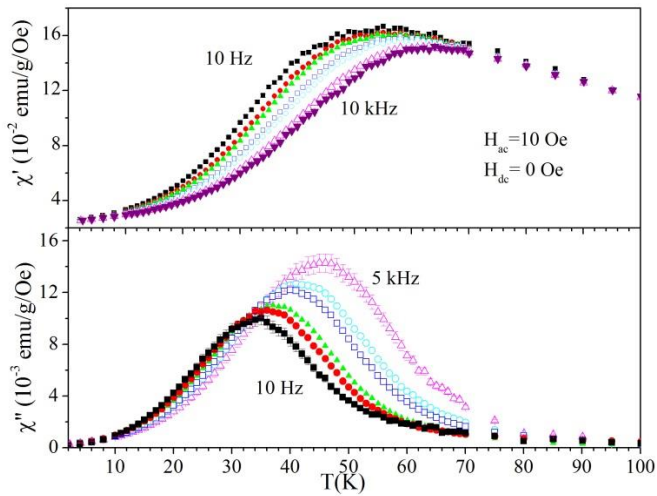

**Fig 7**: Temperature dependence of $\chi'$ (top) and $\chi''$ (bottom) measured at $f_m$= 10, 50, 100, 500, 1000, 5000, and 10000 Hz. The $\chi''$ data at 10 kHz (not shown) is very noisy. The experimental

uncertainties in χ′ are within the size of the data points. For χ″ we are showing the typical uncertainties for the lowest and highest frequencies.

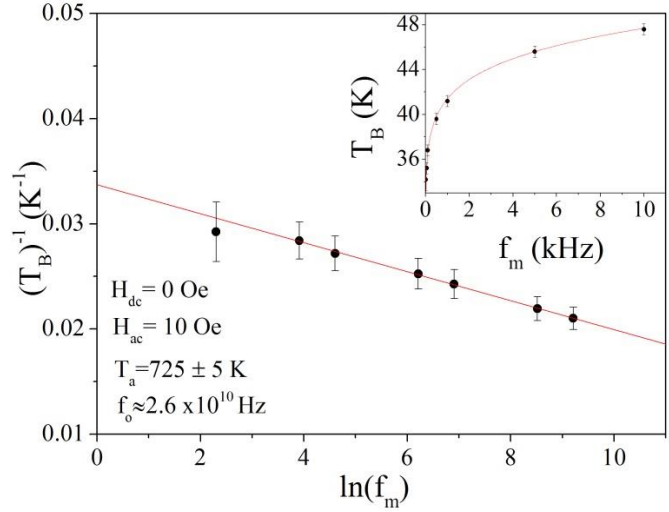

**Fig. 8**: Plot of the inverse blocking temperature vs. ln(f$_m$) following Eq. 2. Linear fit to the data yields T$_a$= K$_a$V/k$_B$=725 K and $f_o$= 2.6 x 10$^{10}$ Hz of Eq. 2. Inset shows the plot of T$_B$ vs. f$_m$ with the solid curve plotted using Eq. 3 and the above magnitudes of T$_a$, $f_o$ and T$_o$ = 0 K.

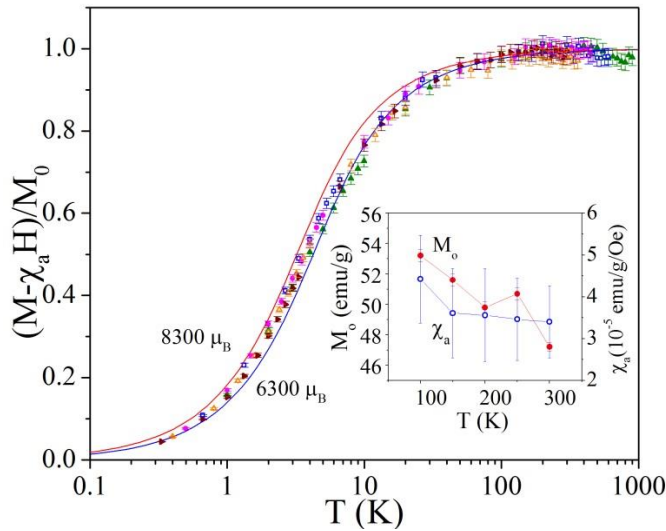

**Fig. 9**: Fit of the M vs. H data of Fig. 6 at T= 100 K, 150 K, 200 K, 250K and 300 K to the modified Langevin function of Eq. 5 with the determined parameters listed in Table1. The two

solid lines represent the simulated fits to Eq. 5 with $\mu_p$= 6300 $\mu_B$ and 8300 $\mu_B$. The inset shows the temperature dependence of $M_o$ and $\chi_a$ determined from the fit with the lines connecting the data points shown for visual clarity.

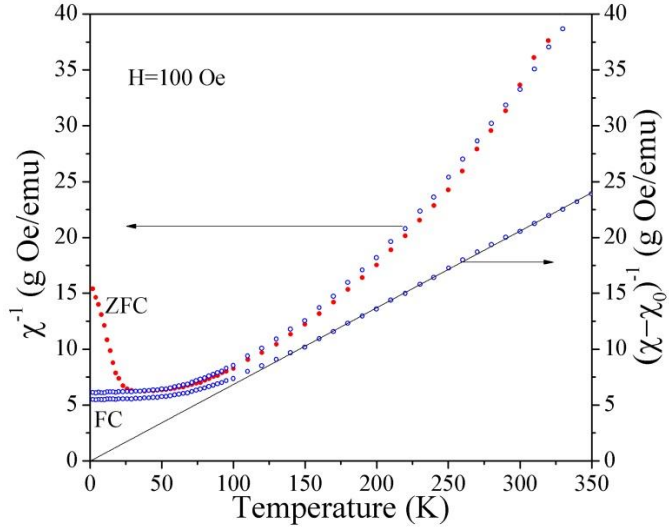

**Fig 10:** Following Eq. 7, the plots $\chi^{-1}$ vs T and $(\chi-\chi_o)^{-1}$ vs T are shown with the slope of the solid line yielding the constant C of Eq. 7.

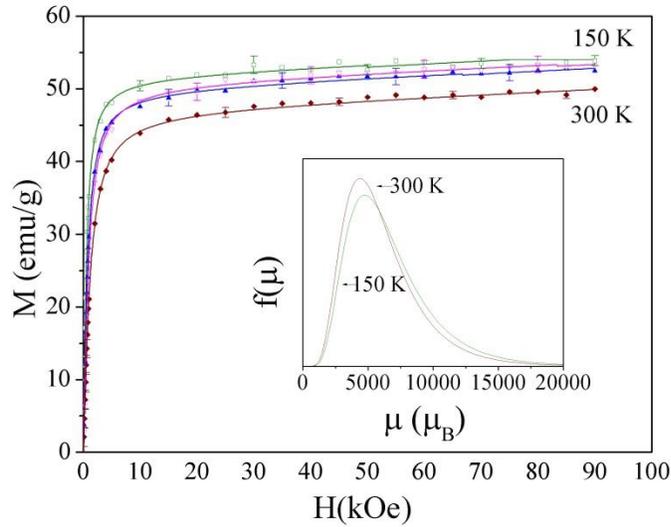

**Fig. 11**: The M vs. H data at T= 150, 200, 250, and 300 K are shown with the solid lines as fits to Eq. 9 and the parameters determined from the fits shown in Table 2. The inset shows the typical distribution functions at 150 K and 300 K determined from the fits. Typical experimental uncertainties are shown for select data points.

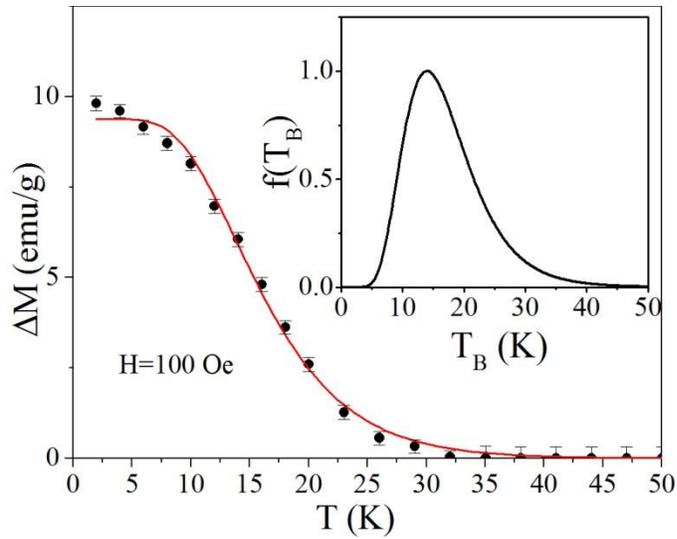

**Fig. 12** The temperature dependence of the difference magnetization $\Delta M = M(FC) - M(ZFC)$ at $H = 100$ Oe determined from the data of Fig. 5 with the solid line is obtained from the fit to Eq. 11. The inset shows the derived plot $f(T_B)$ vs. $T_B$ from the fit.